# A fungal mycelium templates the growth of aragonite needles

Achiya Livne[a], Sylwia Carolina Mijowska[a], Iryna Polishchuk[a], Wilson Mashikoane[b], Alexander Katsman[a] and Boaz Pokroy[a]*

[a]*Department of Materials Science and Engineering, Technion – Israel Institute of Technology, Technion City, 3200003 Haifa, Israel. E-mail: bpokroy@technion.ac.il*

[b]*European Synchrotron Radiation Facility (ESRF), Grenoble, France*

Fungi live within diverse environments and survive well under extreme conditions that are usually beyond the tolerance of most other organisms. In different environments fungi are known to induce precipitation of a wide range of minerals. Various species of fungi have been shown to facilitate calcium carbonate mineralization. Here, inspired by examples of needle-fiber calcite formed *via* fungus-induced biomineralization typically observed in soils and sediments, we utilized inactivated fungus to synthetically induce precipitation of $CaCO_3$ needles. To our knowledge, the feasibility of growing aragonitic needles within fungal mycelium *in vitr*o has not been previously demonstrated. The needles we obtained were curved, displayed hexagonal facets, and demonstrated high-aspect ratios close to 60. The size and shape of these synthetic needles matched those of the mycelium of the natural fungus. Utilizing high-resolution characterization techniques, we studied the morphology and the micro- and nanostructures of the aragonitic needles. Our findings showed that even inactivated fungal mycelium, if present in the crystallization environment, can serve as a template for the formation of high-aspect ratio fibers and can stabilize metastable polymorphs.

# 1 Introduction

Biomineralization is a widespread phenomenon through which minerals are formed by living organisms[1,2]. Biomineralization is observed in all five kingdoms of life, with a large variety of biominerals formed specifically in the Fungi kingdom[3–9]. Since the biominerals produced by a specific fungus differ according to the environment, they are classified as biologically induced[6]. Fungi exist in diverse environments and can survive well even in extreme conditions in which other organisms cannot[7]. They play an important role in the formation of soil, and have a significant role in geochemical processes which shape our planet for millions of years[3,6,7]. An interesting mineral attributed to biomineralization of fungi is needle-fiber calcite. First discovered and named 'moonmilk' in 1555[10], its origin has been debated over the last 50 years. Whereas in the early publications its origin was considered to be inorganic[11], the currently common hypothesis—supported by the size, diameter and branching of the needles[12–15]—is that it originates from mineralization in fungal mycelium. This hypothesis has not been validated, however, since needle-fiber calcite has never been reproduced *in vitro* and has been found in nature only in its final form.

There are several studies in which $CaCO_3$ minerals have been grown synthetically in the presence of fungi. The resulting precipitates were composed mainly of calcite with some vaterite[16]. No documentation of aragonite precipitation in the presence of fungi is available, although this biomineral is common in other organisms[17,18]. Aragonite crystals usually grow with needle morphology along the <111> crystallographic direction. Over the years, various synthetic routes have been developed to control the polymorph of the precipitated

$CaCO_3$. Parameters such as reactant concentration[19], temperature[20], high Mg/Ca ratio[21] and surfactant usage[22] can lead to the formation of aragonite and influence the aspect ratio of the needles[23]. More complex methods, such as growing aragonite on single-crystal substrates[24] or *via* confinement in small membranes[25], offer better control over polymorph selectivity and morphology. In the plastics and the paper industries, $CaCO_3$ is used as a filler to enhance mechanical properties and to whiten the final products[26]. In the oil industry $CaCO_3$ is added as a bridging agent to the drilling fluids to increase density[27]. In all of these industrial cases there is a demand for stronger and denser materials, which would favor aragonite needles over calcite needles[28,29].

In this study, we present a novel bioinspired approach to the synthesis of high-aspect ratio, curved aragonite needles with pseudo-hexagonal morphology. We employed the principles of biomineralization of fungi to grow aragonite *in vitro*. We utilized mycelium, the flexible and branched high surface-area structure of the fungal cells, as a template for aragonite precipitation. Since the crystallization process related to needle-fiber calcite is known to progress slowly at ambient temperature, we employed hydrothermal conditions to accelerate it. We compared the morphology of the synthesized needles with that of the fungal hyphae (tubular cells). Based on the hypothesis that if the crystals grow within these tubular cells which serve as a template, their morphologies would be akin. To further investigate our hypothesis, we aimed to verify the role of the hyphal morphology in needle growth by destroying the hyphal cells *via* sonication prior to growing the crystals.

## 2  Results & discussion

The first stage of this research was to culture the *Trametes versicolor* fungus (see Experimental section). In brief, *T. versicolor* was grown in the form of spheres (Fig. S1), which were then inactivated and sterilized by washing in ethanol and soaking them in it for 24 hours in order to inactivate and sterilize them. Next, *T. versicolor* spheres were added to 50 ml of 0.015 M $CaCl_2$ solution, to which 50 ml of 0.015 M $Na_2CO_3$ was then added. The mixture was poured into a beaker and stirred for 10 minutes, and then immediately transferred into an autoclave preheated to 134 °C. As a first control experiment the same procedure was performed, but in the absence of fungus.

As an outcome of the synthesis that occurred after autoclaving of the *T. versicolor* mixture for 2 hours at 2.5 bars, we observed curved needles exhibiting pseudo-hexagonal morphologies (Fig. 1a). In the control sample, where no fungus was present, only rhombohedral crystals typical for calcite were formed (Fig. 1b). Needles formed in the presence of the fungus were 1.5−5 μm thick, and up to 200 μm long. Their aspect ratio was at least 1:40 and reached >1:100. Branching of needles (Fig. 1a) resembling mycelium structure supported our hypothesis of their fungal origin. Hyphal diameter ranged from 1.5 to 5 μm and corresponded to needle thickness (Fig. 1c). High-resolution synchrotron powder diffraction, which we performed to determine the phase compositions of the obtained crystals, revealed that in the control experiment only calcite was formed, whereas in the presence of fungus the major phase upon precipitation was aragonite, with small amounts of calcite and vaterite (see Fig. 1e). Quantitative Rietveld refinement performed on the full X-ray diffraction (XRD) patterns revealed that the phase composition in the presence of *T. versicolor* comprised 68.6 wt% of aragonite, 15.5 wt% of calcite and 15.9 wt% of vaterite (see Table S1).

In order to verify that the pseudo-hexagonal fibers comprising the major part of the sample were indeed aragonite, we used a focused ion beam (FIB) to prepare two cross-sectional lamellae, one perpendicular and the other parallel to the needle axis, for analysis by transmission electron microscopy (TEM) (Fig. 2a, d and Fig. S2).

Electron diffraction on TEM confirmed that the needles were composed of aragonite oriented along the <001> direction, while the pseudo-hexagonal facets were along <010> (Fig. 2c and f). Dark field (DF) TEM imaging (Fig. 2b) revealed that the horizontal lamella was composed of several grains with varying contrast, indicating that the needle was not a single crystal but rather a textured polycrystal composed of a number of columnar grains. The various grains were all oriented along the <001> direction, yet were slightly rotated relative to each other along that axis. Selected area electron diffraction (SAED) taken from the areas marked with red circles in Fig. 2b demonstrated the same diffraction pattern, which matched that of aragonite with a [001] zone axis (ZA) (Fig. 2c). SAED results for the longitudinal lamella (Fig. 2d, e and f) showed that the area marked with a red circle in Fig. 2e matched the diffraction pattern of aragonite with a [010] ZA, orthogonal to the <001> direction of growth. DF-TEM imaging of the longitudinal cross section also revealed a polycrystalline nature (Fig. 2e).

Aragonite that forms in geological arenas, or by induced mineralization, typically presents a needle-like morphology, but no preferential orientated and its cross-sectional morphology is not hexagonal [30]. The source of the pseudo-hexagonal morphology of our aragonite orthorhombic lattice was the *(a)-(b)* plane, with a *(b)/(a)* ratio close to $\sqrt{3}$. This is why the morphology of the crystals, when growing along the <001> direction, appeared hexagonal[31]. Since the growth direction of the aragonite columnar grains in our case was <001> the horizontal cross section was parallel to the *(a)-(b)* plane, and therefore displayed pseudo-hexagonal morphology.

Comparison of these results to those of the first control experiment clearly showed that the precipitation of calcium carbonate was strongly affected by the presence or absence of fungal mycelium. This was evident from the different calcium carbonate polymorph induced in its presence, as well as from the high-aspect ratio and the orientation of the needle morphology obtained. Studies by Gadd et al.[4,6,7] on the formation of secondary minerals by fungi showed that fungi are involved in the formation of diverse biominerals and are known to affect the precipitates' morphology and composition[4,6,32]. Fungi were also shown to induce precipitation from solutions containing metal ions such as $Zn^{2+}$, $Pb^{2+}$ and $Cu^{2+}$ [33], and it was therefore suggested that the cell wall of the fungus might serve as a nucleation template for precipitates[34]. In our case, it is reasonable to assume that the aragonitic needle formation was indeed initiated in a similar manner *via* sorption of $Ca^{2+}$ ions on the fungus cell wall. We should stress here that the morphology of the obtained needles resembles that of the needle-fiber calcite found in nature, whose origin is still under debate but has been suggested to derive from fungi-assisted mineralization. To further investigate whether the growth of the aragonitic needles had indeed occurred within the fungal mycelium, we performed a second control experiment in which we exposed the inactivated fungus to high-frequency ultrasound waves for 30 minutes, with resulting disruption of the mycelium cells[35] and destruction of their tubular structure. After sonication of the mycelium in the $CaCl_2$ solution, the same process as that in the original experiment was allowed to continue.

When we performed the precipitation experiment in the presence of sonified mycelium, all of the resulting precipitates were rhombohedrally shaped, as in the first control (Fig. 1d). Furthermore, no needles were found, and synchrotron XRD combined with quantitative Rietveld analysis revealed that the major phase was calcite (94.3 wt%) and the rest was vaterite (5.7 wt%) (see Fig. S3). These findings confirmed the cardinal role of fungal hyphae in the templated formation of aragonite needles.

To further support the conclusion that growth of the aragonitic fibers was templated within the confinement of the fungal hyphae, we compared the actual morphology (diameter and length) of these cells with those of the aragonitic fibers formed. To this end, we subjected the fungal hyphae to crytical point drying in order to maintain their biological morphology without distortions, and then coated them with alumina *via* atomic layer deposition (ALD). The latter procedure added mechanical strength to the cells without changing their morphology, and allowed imaging by scanning electron microscopy (SEM) similar to that performed on biofilms[36]. Using ALD, we were able to deposit thin films with conformal coatings of nanometric thickness, even on objects with aspect ratios of up to 1:1000. The morphology of the coated fungal hyphae was indeed similar to that of the aragonitic fibers with respect to length, diameter, and even branching (Fig. 1c). These results clearly indicated that the aragonite grows within the fungal hyphae tubes, thus leading to the formation of needles with high-aspect ratios. We noticed that the crystallization which occurred under the conditions used here proceeded *via* an amorphous phase. This was also demonstrated previously, when it was reported that fast mixing of ….M is undertaken. It seems likely that the amorphous precursor, which is stable only for a short time because it lacks an additional stabilizer (such as magnesium or soluble organics), becomes stabilized within the confinement of the mycelium microtubes. It was clearly shown elsewhere that microconfinement stabilizes amorphous calcium carbonate (ACC) [25]. In this study the crystallization occurred under the hydrothermal conditions in which it was immediately placed. In our first control experiment, where no mycelium was present, the ACC was not stabilized and crystallized very rapidly to calcite. As to why our experimental setup yielded aragonite, we propose a synergistic effect of confinement, organic templating and temperature. Confinement within a synthetic membrane has been shown to induce aragonite, but in that case the pore diameter was only ≈ 25 nm[25]. Here we added a third control experiment performed under hydrothermal conditions, in which the fungal hyphae were replaced by a synthetic membrane with micron-sized pores (of the same order as the mycelium pore diameter). Here too there was no formation of aragonite, but only of calcite. Taken together, the different control experiments clearly demonstrated that crystallization of aragonite or needle-like crystals cannot be induced solely by the effect of temperature, micron-sized confinement, or disrupted mycelium, but only by the synergistic action of all of these effects.

Whether the ACC crystallization occurred by direct transformation from ACC or *via* dissolution-reprecipitation is unknown. We suggest that a hydrated ACC precursor, when stabilized within confined mycelium microtubes, may indeed transform to different calcium carbonate polymorphs *via* the dissolution-reprecipitation mechanism. Both calcite and aragonite may nucleate heterogeneously on the inner walls of the organic microtubes. Calcite is thermodynamically more stable than aragonite, but the kinetics of nucleation (incubation time and nucleation rates) for calcite and aragonite are concurrent and can differ markedly at different temperatures. At room temperature calcite nucleates faster, while increasing the temperature to 100 °C favors aragonite nucleation [ref]. As shown by Munemoto and Fukushi [ref], the incubation time for aragonite nucleation, $\tau_A$, from dissolved monohydrocalcite at temperatures between 10 and 50 °C can be approximated by the dependence:

$$ln\,\tau_A = \frac{13000}{T} - 32, \qquad (1)$$

which corresponds to an apparent nucleation activation energy of ≈ 108 kJ/mol. Dissolved monohydrocalcite can be considered as hydrated ACC. At room temperature, from eq. (1) we obtain $\tau_A \approx 2 \cdot 10^5$ seconds. Estimation of incubation time at 100 °C from eq. (1) yields ≈ 17 seconds, while at 134°C (as in our experiments) it decreases strongly to ≈ 1 second. It should be noted that the formation of calcite in this study was inhibited by the addition of Mg-stabilized ACC.

The apparent activation energy of nucleation is a complex parameter that includes a thermodynamic barrier (work done in formation) of the critical nucleus) and an effective kinetic barrier (arising from desolvation) of solute ions, activation energy for the attachment of ions to the critical nucleus, and structural rearrangements). Kawano et al. [ref] constructed precipitation diagrams for the three calcium carbonate polymorphs (calcite, aragonite and vaterite) showing which of these phases is the first to precipitate from a solution under a given condition. To explain why aragonite formation was favored at 50−80 °C, they suggested that the surface energy of aragonite decreases with temperature according to the relation $\gamma_A = \gamma_{A0} - a(T - 298)$, where $\gamma_{A0} = 150$ mJ/m$^2$ and $a = 8$ mJ·m$^{-2}$K$^{-1}$, while surface energies of calcite and vaterite remain constant with temperature. However, such an assumption is difficult to justify. Moreover, extensive calcite formation (while there was no aragonite formation) in our experiments with the sonified mycelium at 134°C contradicted this hypothesis. In our opinion, the formation of calcite and the absence of aragonite during high-temperature crystallization in the absence of tubular mycelium cells can be explained by the absence of an intermediate stabilized and hydrated ACC. In such a case, calcite is crystallized directly from the supersaturated solution. On the other hand, nucleation of crystalline calcium carbonate is known to be heterogeneous, and to be controlled by the electrostatic interaction between the hydrated calcium ions and the substrate (ref). Our calculations supporting this suggestion can be seen in Supplementary Materials. Based on these calculations it was apparent that the nucleation of aragonite in our case was kinetically much more favorable, despite the higher thermodynamic stability of calcite.

## 3 Conclusion

Here we demonstrated for the first time the feasibility of growing aragonite needles within inactivated cells of fungal hyphae *in vitro*. Our results clearly showed that aragonite needles are formed by the synergistic effect of confinement, organic templating and crystallization temperature. Notably, this study demonstrated the potentially important role played by mycelium organics in fungus-induced biomineralization, and it also strengthened the notion that needle-fiber calcite originates from fungi. Further investigation may be helpful in developing a novel approach to control crystal formation that is both simple and cheap. The present study, our first in this field of research, was performed with $CaCO_3$ minerals, but the findings might possibly be extended to other engineering materials as well. This may lead in turn to utilization of fungal cells or fungi-inspired surfaces as organic templates for crystal growth.

## 4 Experimental

### 4.1 Growth of fungus

The fungus *Treametes versicolor* (kindly supplied in its original form by Prof. S.P. Wasser) from the Department of Evolutionary and Environmental Biology of Haifa University was grown on agar substrate. Standard malt extract agar (30 g malt extract, 10 g peptone and 15 g agar per 1 L of deionized (DI) water was sterilized by autoclaving at 120 °C for 45 min, poured into 90-mm Petri dishes and cooled to room temperature while covered with lids inside a sterile biological hood. A piece (1×1 cm) of living fungus on agar was attached to the center of each Petri dish, the lids were re-sterilized by flame, and the Petri dishes were sealed with parafilm and kept in an incubator at 25 °C for 7 days, by which time the whole agar surface in each dish was covered by the fungus. A standard growth solution (15 g D-glucose, 2.5 g peptone, 3 g yeast extract, 1 g $KH_2PO_4$, 0.2 g $K_2HPO_4$ and 0.5 g $MgSO_4$ per 1 L of DI water) was prepared in 50-ml Erlenmeyer flasks, and after full dissolution the pH was set to 5.5 using NaOH and HCl. Each flask was closed with a cellulose stopper, covered with aluminum foil, autoclaved at 120 °C for 45 min, and then placed under the biological hood and allowed to cool to room temperature. From each Petri dish, five slices of *T. versicolor* (an area of 5×1 mm at the depth of the Petri dish) were transferred to an Erlenmeyer flask, and the neck and cellulose stopper of each flask were sterilized by flame. The system was sealed and kept for 3 weeks in an incubator at 25 °C, during which time the flasks were continuously shaken at 120 rpm on an orbital shaker. By the end of that time the fungus had grown into spheres of ≈ 1 cm diameter.

**4.2 Establishment of hydrothermal conditions**

Each *T. versicolor* sphere was washed in AR-grade ethanol and soaked in it for 24 h to inactivate the fungus and remove any organic compounds. Each sphere was then transferred into a 150-ml beaker containing a solution of 50 mL $CaCl_2$ (0.015 M) and was stirred for 10 min. To this was added a 50-ml solution of $Na_2CO_3$ (0.015 M) from a 100-ml beaker, and the mixture was immediately placed in an autoclave preheated to 134 °C, and kept there for 2 h at 134 °C and 2.5 bars. The solution was then filtered through Whatman paper no. 5 and air-dried for 24 h. A first control experiment was performed under identical hydrothermal conditions at all stages of preparation, but with no fungus added to the system.

To verify that the aragonite needle's morphology is determined by the fungal tubes, we produced a sonicated sample by exposing mycelium's tubular cells to high-frequency ultrasound waves for 30 min in order to explode the cells and thereby destroying the mycelium's tubular structure. After sonication, we continued to subject it to the same process as that used for the original (unsonicated) sample.

Fitting of the needles to cells of the fungus required the use of fungal cells in their original form, not cells that had collapsed under vacuum. We used critical point drying and atomic layer deposition (ALD) to achieve a well-preserved mycelium structure. The *T. versicolor* spheres were moved to ethanol AR and shaken on an orbital shaker at 120 rpm for 3 days. The ethanol was then replaced with fresh ethanol, and shaking was continued for 4 more days. The clean and water-dry fungus was then placed in the chamber of a critical point dryer, in which $CO_2$ under pressure was substituted for ethanol in several cycles. Upon full substitution, the temperature and pressure reached the $CO_2$ critical point, and the $CO_2$ was vaporized. The critical point-dried fungus was then subjected to $Al_2O_3$ deposition in an ALD chamber (SUNALE$^{TM}$ 200 reactor, Picosun, Finland). Since we were using organic materials as a template we needed to work at relatively low temperatures, and accordingly kept the growth temperature at 80 °C. Trimethylaluminum (TMA; 97% pure) was used as an aluminum precursor and water vapor as the oxidizer. To maintain the reactor pressure at 10 hPa we used a $N_2$ carrier gas flow (99.999% pure) of 300 cm$^3$min$^{-1}$. The sources of TMA and $H_2O$ were kept at room temperature. $N_2$ flow rates during TMA/$H_2O$ fill and purge periods were 100 cm$^3$min$^{-1}$. Stop flow mode consisting of 350 cycles was used for the run.

**4.3 Sample preparation**

The solution was filtered through Whatman filter paper no. 5. The powder was air-dried for 24 h. We produced SEM samples by placing powder onto conductive carbon tape on a SEM sample holder. Samples were carbon-coated twice. Powders for X-ray diffraction (XRD) were not grinded. TEM samples were produced by the use of a FIB. From the SEM samples we cut two lamellar cross sections, one horizontal and the other longitudinal.

**4.4 Characterization**

SEM images were obtained with a Zeiss Ultra plus high-resolution SEM (HRSEM). Energy-dispersive X-ray spectroscopy (EDS) measurements were obtained using the same SEM at 10 keV. XRD measurements were carried out at the ESRF (European Synchrotron

Radiation Facility) in Grenoble, dedicated beamline ID22. Wavelength was 0.4000 Å, with a scan rate of 4 degrees. min$^{-1}$. For TEM measurements on the horizontal cross section we used the FEI Titan Themis Cubed G2 at 300 keV, and for TEM measurements on the longitudinal cross section we used the FEI Tecnai T20 at 200 keV.

**Acknowledgements**

We thank Prof. S.P. Wasser from the Department of Evolutionary and Environmental Biology, Haifa University, for donating the fungus species used for these experiments and teaching us how to grow it in our lab. The diffraction experiments were performed on beamline ID22 at the European Synchrotron Radiation Facility (ESRF), Grenoble, France.

**Notes and references**

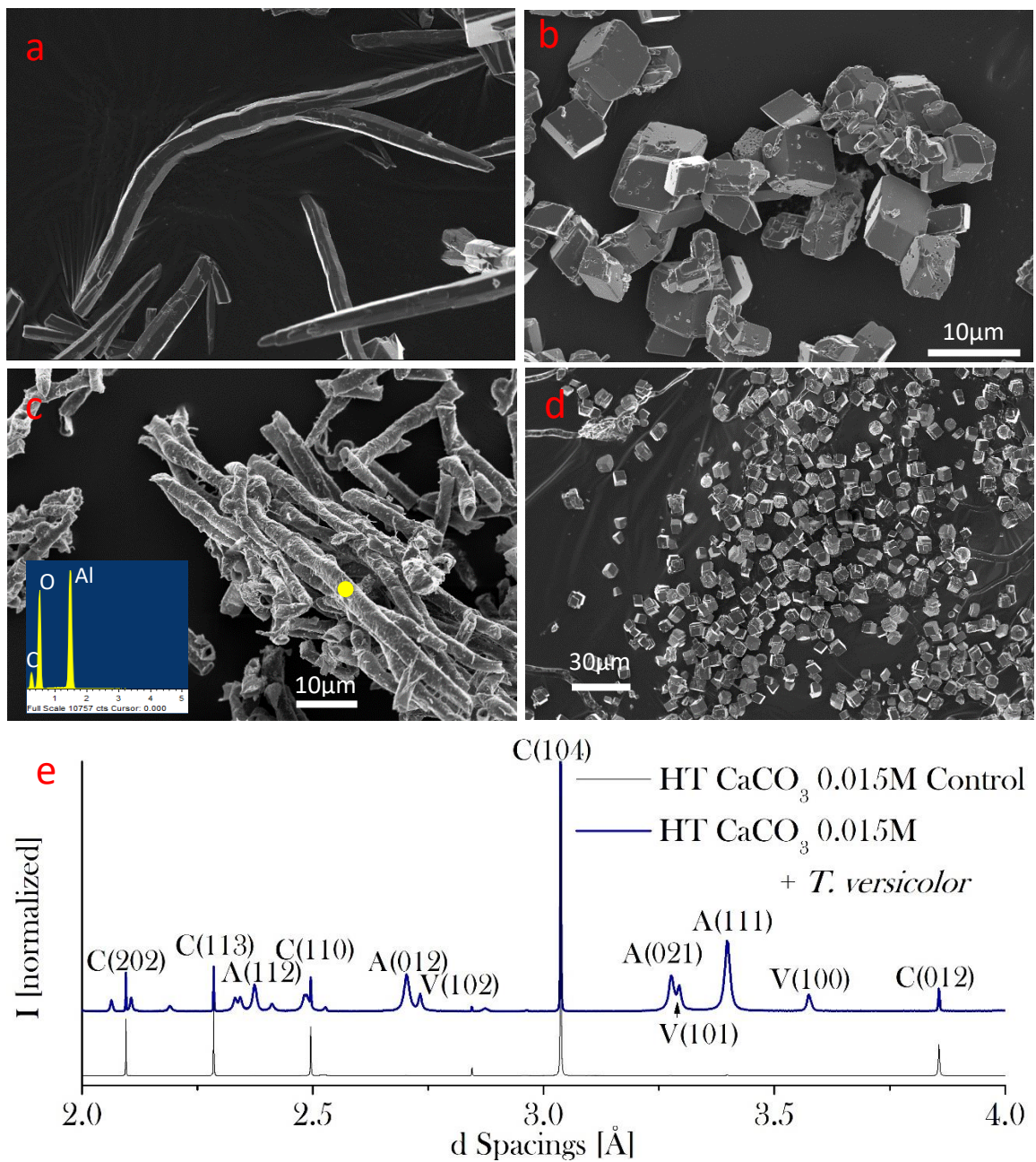

**Fig. 1.** Characterization of CaCO$_3$ crystals mineralized with and without fungus. (a) SEM image of control crystals obtained under hydrothermal conditions in the absence of fungus. Rhombohedral calcite crystals are the main precipitate. (b) SEM image of the crystals obtained in the presence of *T. versicolor* fungus under the same conditions. Curved aragonite needles with hexagonal facets are formed. Needle thickness ranges from 1.5 to 5 μm. (c) SEM image of *T. versicolor* mycelium coated with alumina. Hyphal thickness ranges from 1.5 to 5 μm. Inset shows the EDS spectrum taken from the area marked in yellow and confirming that the tubes are covered by Al$_2$O$_3$. (d) SEM image of crystals obtained when the fungus was sonicated prior to the synthesis. No precipitation of aragonite needles is observed. (e) XRD patterns of the control sample (without fungus) and of the sample grown in the presence of fungus. Control crystals show a pattern of the pure calcite (black line). In the presence of fungus aragonite and vaterite phases are observed (blue line).

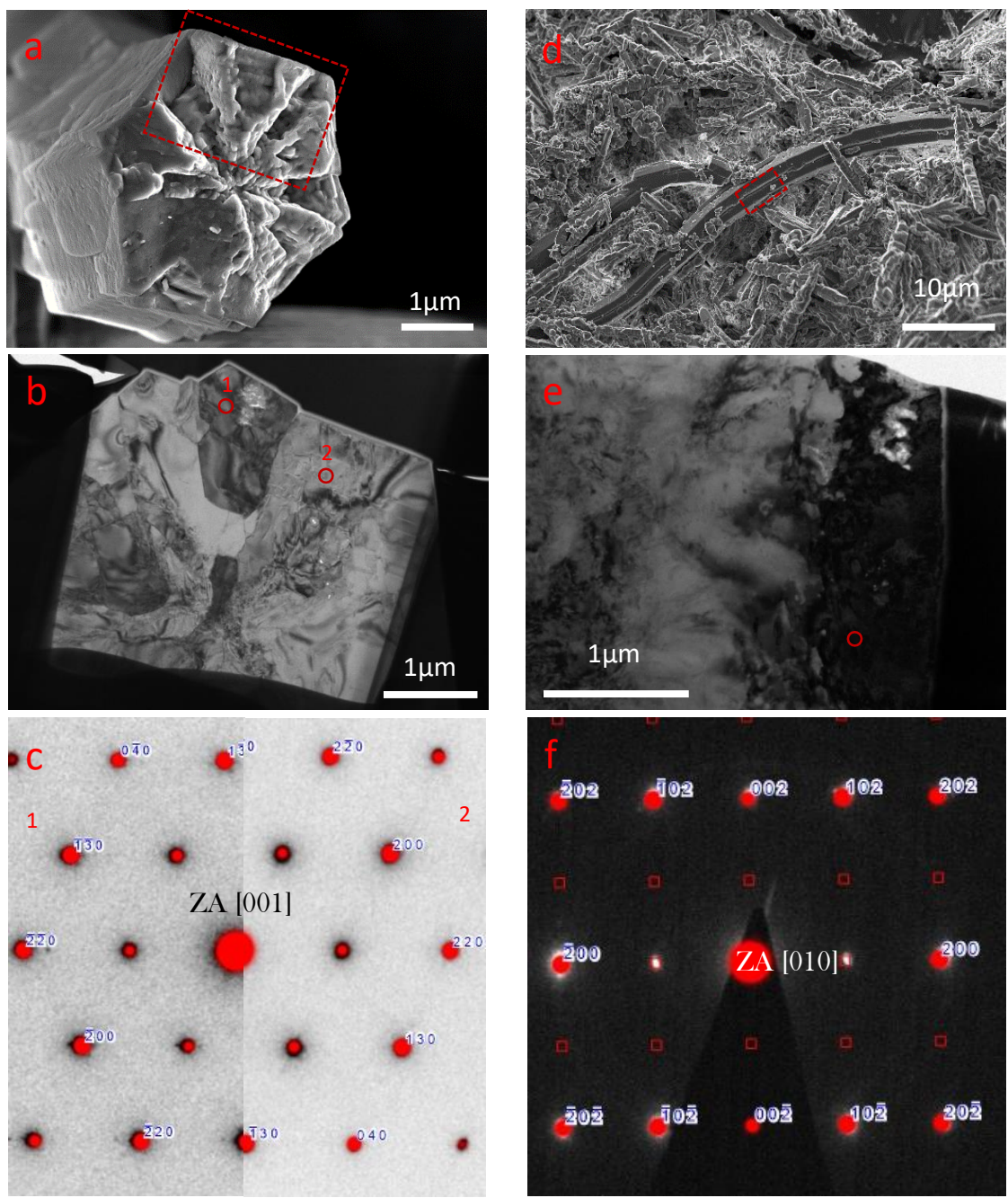

**Fig. 2**. TEM results obtained from horizontal and from longitudinal lamellae of a needle. (a) SEM image shows a needle's hexagonal shape. The red rectangle marks a horizontal cross section cut with a FIB. (b) Dark-field TEM image of the horizontal lamella. The varying contrast indicates that the needle is polycrystalline. Red circles mark areas of selected area electron diffraction (SAED). (c) SAED of the horizontal cross section from both 1 and 2 (left and right respectively), fits to aragonite with zone axis [001]. (d) SEM image of a curved hexagonal needle; the red rectangle marks the longitudinal cross section. (e) Dark-field TEM image of the longitudinal lamella. The varying contrast indicates that it is not a single crystal. Red circles mark the areas of SAED. (f) SAED of the longitudinal cross section. The diffraction matches that of aragonite with zone axis [010].